\documentclass[]{aa}
\usepackage{txfonts}
\usepackage{natbib}
\usepackage{graphicx}
\nonstopmode

\begin{document}

\title{Dust and gas in the central region of NGC 1316 (Fornax A) - Its origin and nature
 \thanks{Based on observations taken at the European Southern Observatory, Cerro Paranal, Chile, under the programme 92.B-0289}}
    
\subtitle{}

\author{
T. Richtler     \inst{1} 
\and 
M. Hilker \inst{2}
\and
E. Iodice \inst{3}
}
\offprints{T. Richtler}
\institute{
Departamento de Astronom\'{\i}a,
Universidad de Concepci\'on,
Concepci\'on, Chile;
tom@astro-udec.cl
\and
European Southern Observatory,
Karl-Schwarzschild-Str.2,:
85748 Garching,
Germany
\and
INAF-Astronomical Observatory of Capodimonte, via Moiariello 16, Naples, I-80131, Italy}
\date{Received  / Accepted }

\abstract
 {
 The  early-type galaxy NGC 1316, associated with the radio source  Fornax A,  hosts about $10^7 M_\odot$ of dust within a  central radius of  5 kpc. These prominent
 dust structures  are believed to have an external origin, which is also a popular interpretation for other dusty early-type galaxies. 
  }
 {Our aim is to understand the nature of ionised gas and dust in NGC 1316 and  to offer
an interpretation for the origin of the dust. }
 {We use archival Hubble Space Telescope/Advanced Camera for Surveys  data to construct colour maps that delineate the dust pattern in detail, and we compare these  data with  maps
 constructed with data from the  Multi Unit Spectroscopic Explorer (MUSE) instrument of the Very Large Telescope at the European Southern Observatory.  
  We use the tool PyParadise to fit the
 stellar population. We use the residual emission lines and the residual interstellar absorption NaI D-lines, and  we measure line strengths, the velocity field, and the
 velocity dispersion field. 
  }
 {The emission lines resemble low-ionisation nuclear emission-line region (LINER) lines, with [NII] being the strongest line everywhere. 
 There is a striking match between the dust structures, ionised gas, and atomic gas distributions, the last of which  is manifested by interstellar absorption residuals of the stellar NaI D-lines. In the dust-free regions, the interstellar NaI D-lines
 appear in emission, which is indicative of a galactic wind. 
 At the very centre, a bipolar velocity field of the ionised gas is observed, which we interpret as an outflow.  
 We identify a strongly inclined  gaseous dusty disc  along the major axis of NGC1316.                            
  A   straight  beam of ionised gas with a length of about 4 kpc emanates from the centre.   }
 {
       Our findings are strongly suggestive of a dusty outflow that is curved  along the line-of-sight. Nuclear outflows may be  important dust-producing  machines in galaxies.
 (Abridged for arXiv)
 }
   
\titlerunning{Dust and more in the centre of NGC1316}

\maketitle

\section{Introduction}
  
The origin of dust in early-type galaxies  after some 30 years  of cartography  \citep{ebneter85,goudfrooij94,vandokkum95,tran01,lauer05,patil07,finkelman10}  is 
still a matter of discussion and involves a wide range of physical processes. The best understood dust producers are
probably supernovae (SNe) \citep{cherchneff17} and asymptotic giant branch (AGB) stars \citep{dellagli17}. If young stellar populations are lacking, as is the case in early-type galaxies, the 
dust can be attributed to infall or to other processes such as metal-accretion in the gas phase   \citep{hirashita99}. 
  A concise introduction to the literature on dust in early-type galaxies is available in \citet{hirashita15} and \citet{hirashita17}.

One of the most conspicuous dust structures in the nearby Universe
is found in NGC 1316 (Fornax A), which is located on the outskirts of the Fornax cluster.
The dust structures in NGC 1316 are so striking that H. Shapley mistook them for plate defects on old photographic plates \citep{schweizer80, hodge75}. 
NGC1316 has the reputation of being a merger remnant. Many shells and ripples are reminiscent of previous galaxy interactions and the infall of dwarf galaxies \citep{mackie98,richtler12a,iodice17}. 

The dust,  with a total mass of about  $1-2 \times10^7$ $M_\odot$ \citep{remy14},
 has been attributed to infall in all investigations to date. In a comprehensive work on interstellar matter in NGC 1316, \citet{lanz10} proposed  the infall of an Sc- or Sm-galaxy, in view of the  dust-to-gas ratio
of a putative merger. 

However, a simple look at the dust morphology revealed by Hubble Space Telescope/Advanced Camera for Surveys (HST/ACS) images raises doubts about the dust being brought in by infalling galaxies (see Fig. \ref{fig:largescale}). The radial symmetry is 
striking, as already noted by \citet{searle65} and \citet{schweizer80}. The outer boundary is a well-defined circle, the two main dust axes
point away from the centre, and there are many radial structures among the dusty filaments.  Moreover, if the dust has  its origin in AGB or SN winds, these young
populations should be identifiable; however, the youngest populations have ages of about 1-2 Gyr, with only small contributions from slightly younger populations \citep{richtler12a}.

 \citet{schweizer80} detected optical emission lines that indicated the existence of a fast rotating central disc. 
He also noted the weakness of  H$\alpha$ with respect to [NII], which nowadays is interpreted as post-AGB stars being ionising sources
  (we discuss this further in Sect. \ref{sec:NII}).
The interstellar medium (ISM) of NGC 1316 has been investigated in the mid-infrared (e.g. \citealt{lanz10}, \citealt{duah16})
through observations of molecular gas at mm wavelengths \citep{horellou01,morokuma19}
 and of the HI 21 cm line \citep{horellou01, serra19}. In the optical, new spectroscopic observations with Keck were recently
presented by \citet{morokuma19}.
 In this paper, we will investigate the nature of the ionised gas and
its close relation with the dust structures. We will also confirm the explanation from \citet{morokuma19} for Schweizer's disc. 

We present  observational arguments, based on Multi Unit Spectroscopic Explorer (MUSE) data and HST/ACS images, 
that strongly suggest an internal origin of the major part of the dust within NGC 1316, most probably  in a nuclear dusty wind, which would be a new  dust production or distribution machine for galaxies.
The MUSE data open
 up a plethora of research lines regarding the  analysis of the stellar population, its kinematics and dynamics, and  diagnostics of emission lines.       We will leave a full account of spectral and kinematic MUSE data to future publications.

 Our present scope  is a first compact phenomenological description of  the new picture  and the identification of  particularly intriguing features. 
 These phenomena are: the stunning match of the dust distribution with ionised and neutral gas; the gaseous velocity fields; a nuclear outflow; and some radial jet-like features.
A new diagnostic tool for neutral gas in early-type galaxies is the use of the NaI D1/D2 lines in absorption and emission. 

We adopt the SN type-Ia distance of 17.8 Mpc quoted by \citet{stritzinger10}, corresponding to a distance modulus of $(m-M)_0=31.25$ mag; however, \citet{cantiello13} quote a larger distance of 20.8 Mpc. One arcsecond corresponds to 86.3 pc for a distance of 17.8 Mpc.

 


\section{Observations and reductions}
\label{sec:obs}
NGC 1316 was observed with the MUSE instrument mounted on the Very Large Telescope (VLT) at the European
Southern Observatory (ESO) in Cerro Paranal, Chile, under the programme 94.B-0289 (PI: J. Walcher).  
MUSE is a mosaic of 24 integral field units (IFUs), which covers a field of 1$\times$1 arcmin$^2$ in the wide field mode. The pixel scale is 0.2$\times$0.2 arcsec$^2$. 
 The spectral  resolution varies from R=2000 at 4700 \AA\ to R=4000 at 9300 \AA,\ which are the extremes of the spectral range covered.  The entire observed field in NGC1316 is a mosaic of 12 pointings,
 arranged as 3$\times$4 rectangles and centred on the nucleus. The major axis  has a position angle of 40$^\circ$.  Each of the 12 fields has been
 observed twice with exposure times of 150s each. 
The data were  taken on the nights of 13 December 2014 and 27 December 2014.    



\subsection{Reduction}
The ESO Phase 3 concept offers reduced data products through the ESO science archive. However, by the time this concept became operative, our dataset had been already reduced.
The Phase 3 products are not superior to our original reduction, but we want to call attention  to the fact that they exist.

 The reduction was performed using the pipeline provided by ESO, as described by the pipeline manual version 1.6.2, employing the standard EsoRex recipes. 
 The basic reduction consists of applying {\it muse\_bias} and {\it  muse\_flat} (no correction for dark currents), followed by the wavelength-calibration with {\it  muse\_wavecal}. 
 The line-spread-function was calculated from the arc spectra using {\it muse\_lsf}. For the instrument geometry, we used the tables provided by ESO. Twilight exposures
 were used for the illumination correction, applying {\it muse\_twilight}.   The previous recipes produce tables
 that  were then entered into the recipe {\it muse\_scibasic}, which performs bias subtraction,
 flat field correction, wavelength calibration,  and more. The recipe {\it muse\_scipost} performs flux calibration and calculates the final data cube, or, if desired, calculates fully reduced pixtables that are combined by {\it muse\_exp\_combine} to
 produce a data cube with combined individual exposures. The pipeline also corrects for  telluric absorption features.
 We expected the relative flux uncertainties to be about, or better than,  5\%, which was the overall experience (e.g. \citealt{weilbacher14}).

\subsection{The Paradise software}
This paper deals  with emission lines and interstellar absorption and does not investigate the stellar population. However, to isolate emission lines,  the spectrum of the galaxy light  had to be fitted by population
synthesis  and subtracted.
The best fit is found as a combination of template single stellar population spectra,  which, in our case, are models from \citet{bruzual03}. 
We do that by employing the PyParadise software \citep{husemann16}, which is an extended python version of Paradise. Details of the fitting procedure are explained in the appendix of  \citet{walcher15}.


\subsection{HST images}
We compared the MUSE data with optical HST/ACS images, which    cover the central dust features.  We made use  of the Hubble Legacy archive  and 
 the HST  datasets  $\rm hst\_9409\_01\_acs\_wfc\_f435w\_drz.fits$  and $\rm hst\_9409\_01\_acs\_wfc\_f814w\_drz.fits$  (Proposal number 9409, PI: P. Goudfrooij). These images were taken on 7 March 2003 with
exposure times of 1860s and 4860s for the filters F435W  and F814W, respectively.

\section{The large-scale picture}
In the following, we present and compare the distribution of three components of the ISM: dust, ionised gas, and neutral gas.

\subsection{A central reddening map for NGC1316}
\begin{figure*}[th!]
\begin{center}
\includegraphics[width=1.0\textwidth]{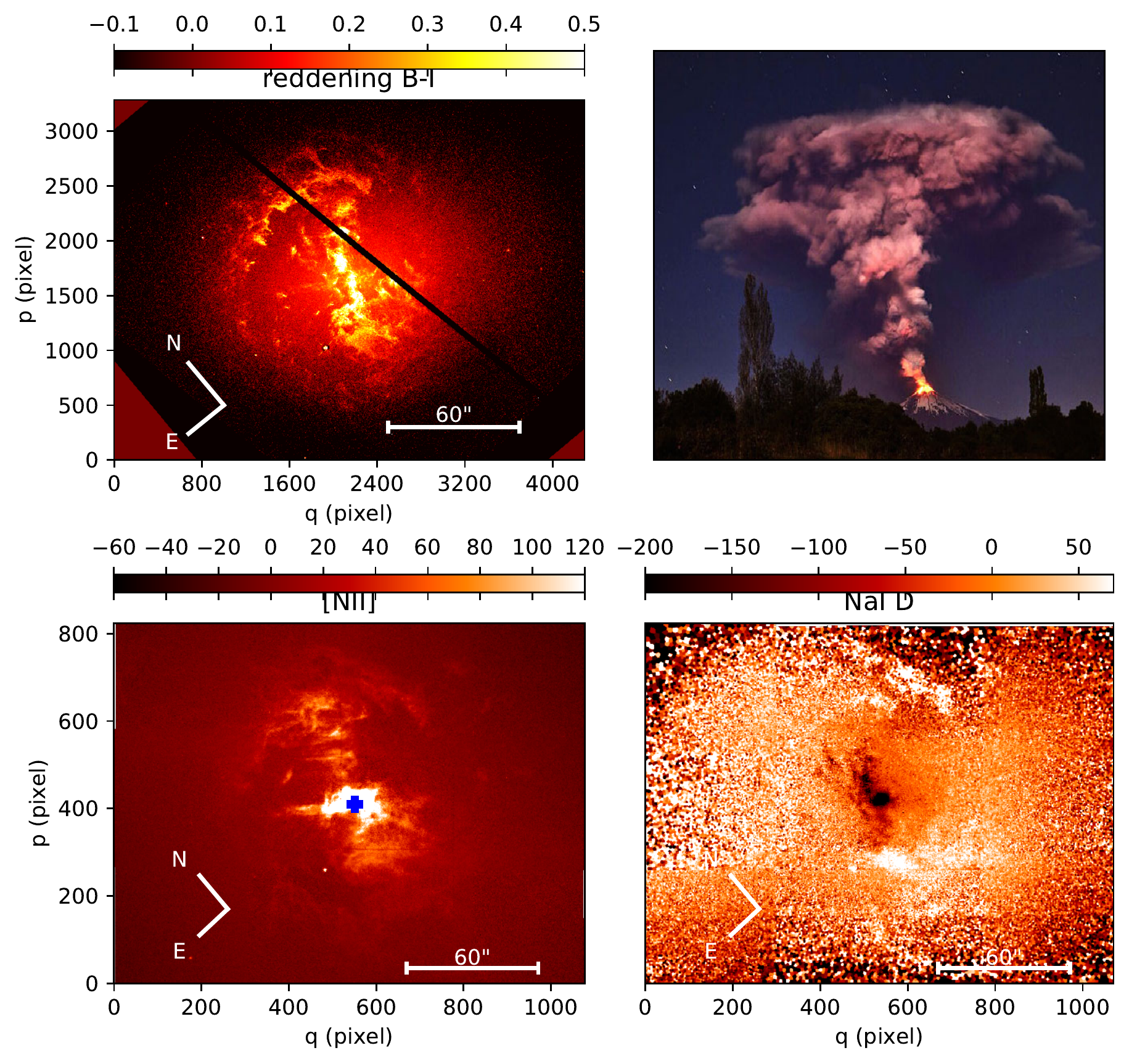} 
\caption{Maps of various components of the ISM in NGC 1316. {\bf Upper-left panel:} Reddening map in B-I based on  HST/ACS exposures, the best way to illustrate the dust distribution. We note the spherical shape and the  axial symmetry with respect to the centre. The galaxy's major axis is aligned with the horizontal axis of this image, as are the other NGC 1316 images in this figure. The black line is the chip gap. {\bf Upper-right panel:}
Eruption of the Villarica volcano in Chile in March 2015. The morphological similarity triggered us to view the nucleus of NGC 1316 as the main source of  dust. {\bf Lower-left panel:}  MUSE mosaic in
the light of the [NII]-line 6583\AA, which is the strongest line everywhere. The ionising sources are plausibly AGB stars. We note the close  resemblance to the dust distribution. Many details in the dust find their counterparts in the [NII]-structures. {\bf Lower-right panel:}
Interstellar absorption line NaI D after Voronoi tessellation and the subtraction of the model galaxy light. Many dust features appear again in this map, demonstrating the quality of this technique as a probe of atomic gas. We note
the line emission in those regions that 
are less obscured by dust. We conclude from the close relation of dust, ionised gas, and atomic gas that gas and dust were never as separated as is normally the case in galaxies: They must
come from the same small volume, most probably a nuclear wind.}
\label{fig:largescale}
\end{center}
\end{figure*}

The main dust structures in NGC 1316 are found within a radius of approximately 1 arcmin (5.2 kpc), which can be covered by one HST/ACS field. The best visualisation of the dust against the very bright galaxy background is
a reddening map  (Fig. \ref{fig:largescale}, upper-left panel).
To facilitate the comparison with  standard reddening laws, we transformed the ACS photometry into Johnson B-I, for which we used the relations given by \citet{sirianni05}. The colour between
the dust features near the centre varies between  2.1 $<$B-I $<$ 2.2, but we cannot exclude the existence of  structureless  dust that causes a small amount
of reddening, which is otherwise invisible.
 We therefore adopted $(B-I)_0 = 2.0$ as the unreddened colour of the central region of NGC 1316 and interpret
all deviations to redder colours as reddening caused by dust.  

 Such a map reveals extremely fine details that cannot be appreciated  in a figure like Fig. \ref{fig:largescale} (upper-left panel), which is needed for comparison with the MUSE mosaic. The central
and axial symmetries are striking (which has been noted above).  The picture of the eruption of the Villarica volcano in Chile (upper-right panel) shows a stunning morphological similarity to the northern part of the NGC1316 dust structures. We do not
stretch the analogy of the physical processes too much, but it was this picture that first triggered the idea to see the dust in NGC 1316 (or part of it) as  being generated in the central region. The dynamics of a volcanic eruption
column are driven by the turbulent mixing of the hot volcanic dust (with grain sizes similar to interstellar dust) with the ambient air resulting in a convective thrust, while the dynamics of dusty nuclear outflows are probably
governed by radiative transfer \citep{hoenig17}. However,  structure formation along a dusty wind plausibly depends, in a complicated manner, on the spectrum of particle velocities as well as on particle interaction with the ambient
medium (which is hot in the case of NGC1316)  and may be scale-free.

\subsection{Morphology and distribution of [NII]}
\label{sec:NII}

\begin{figure}[h]
\begin{center}
\includegraphics[width=0.5\textwidth]{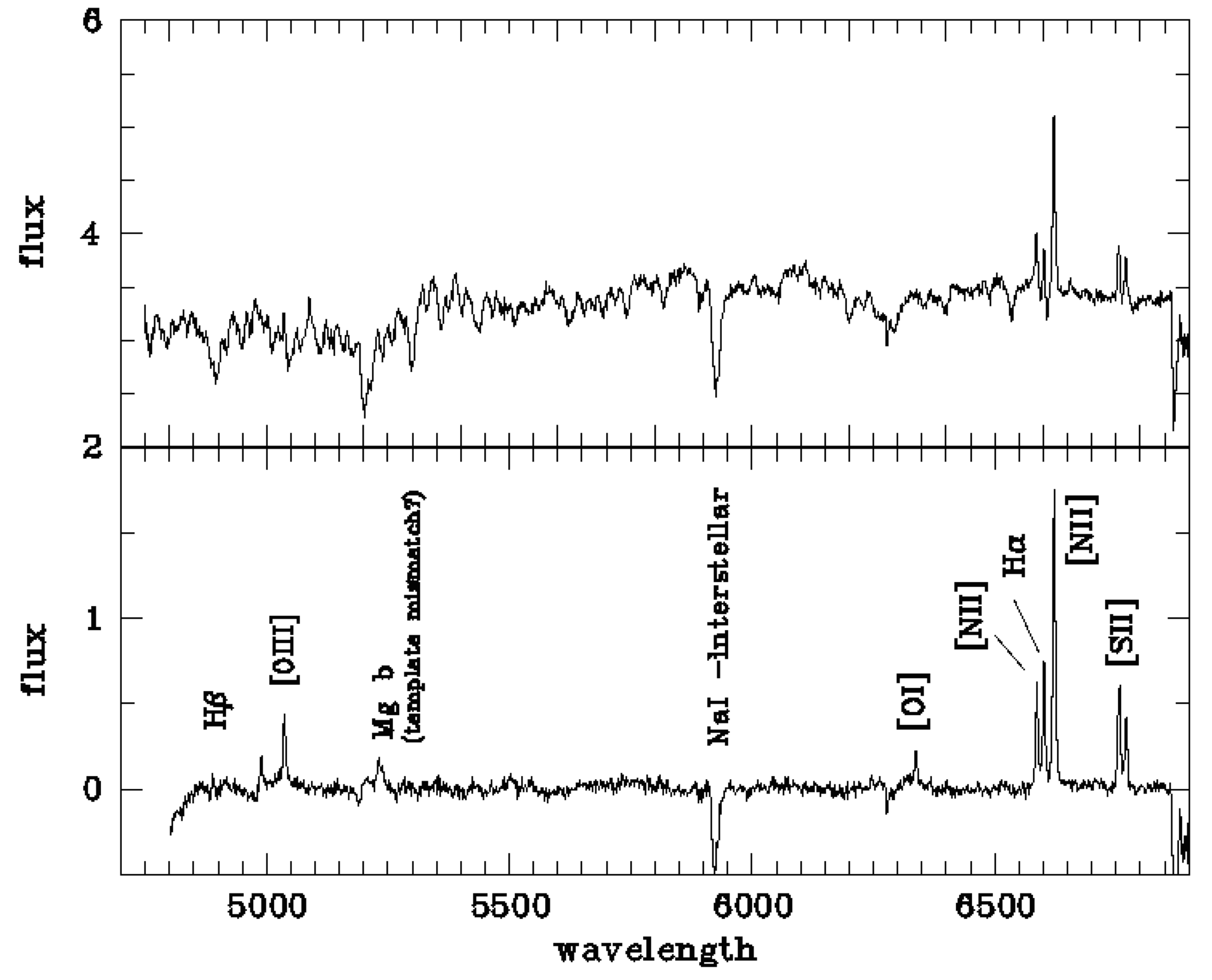} 
\caption{Example spectrum near the centre, {\bf  2.6"N, 0.4"E of the nucleus}. The upper panel shows an old metal-rich galaxy spectrum with a velocity dispersion of about 200 km/s. The stellar
population has been fitted with PyParadise, which {\bf enables us} to subtract the stellar contribution and leave only the emission lines, which are shown in the lower panel.  It is a LINER-like spectrum displaying the typical emission lines, which, in this case, are caused by the radiation field of post-AGB stars. {\bf The residual of the Mg b-band is only visible near the centre and is most probably caused by  template mismatch.} 
  }
\label{fig:spectrum}
\end{center}
\end{figure}

The spectrum shown in Fig. \ref{fig:spectrum} was extracted at 1.5"E and 3.5"N  of the optical centre. It shows a metal-rich, old population superimposed with low-ionisation nuclear emission-line region (LINER) 
emission lines.
The spectral appearance  is quite homogeneous over the  entire field
of detectable lines, as well as far from the nucleus in the outer dust structures where the [NII]-intensity is only a factor of 0.0003 of the central intensity.  This strongly suggests that the ionising sources are very local. Therefore, the  ionising radiation must be  provided by post-AGB stars, as is widely acknowledged for early-type galaxies (e.g. \citealt{cid11,zhang17}). 
The [NII]-line at 6583\AA\  is the strongest line everywhere. We prefer it over H$\alpha$  as a tracer of ionised gas as it has the additional advantage that the local continuum can be easily subtracted in contrast to H$\alpha,$ which
is embedded in a deep absorption trough.

The lower-left panel  in Fig. \ref{fig:largescale} shows  the [NII]-map of the entire MUSE mosaic using the {\it MONTAGE} software. For the mosaic construction, we extracted the [NII]-image  with QFitsView \footnote{QFitsView
was developed at the Max-Planck-Institut for Extraterrestrial Physics by Thomas Ott and Alex Agudo Berbel.}  using the
line map option. The original images were rotated anticlockwise by 40 degrees and then cropped.
The size of the MUSE mosaic is 3.33$\times$2.42 $\prime^2$, corresponding to 17.24$\times$12.53 kpc. 

The overall extension and similarity with the dust structure is striking. A closer view shows that almost all the  [NII]-structures do indeed find their correspondence in the
  dust distribution to  a stunning degree. There are a few exceptions.  Most of the [NII]-luminosity comes from the central region but does not resemble the inner dust structure.  The   [NII]-'cloud' at 
  q$\approx$650, p$\approx$650 has no dusty counterpart but shows NaI D in emission.  Additionally, the few little spots in the western part have no dusty counterparts.

It is clear that this is not the relation of dust and gas that is found in star-forming regions in galaxies. Dust (and molecules) indicate the coldest and densest parts of the ISM in a fractal manner, while the emission gas in HII-regions 
indicates the opposite: lower density, higher temperature, and no star formation.  
We further conclude that there is no, or very little, ionised gas between 
the dust structures because the missing shielding would enhance the ionisation rate and would enhance visibility. 
The close link between the [NII]-emission and the dust also opens up the possibility of studying the dust kinematics  (see Sect. \ref{sec:kinematics}).

\subsubsection{Two radial features:  Disc and jet }
In  the morphology of the [NII]-map (Fig. \ref{fig:largescale}, lower-left panel), two radial (i.e. emerging from the central region) features are striking.  One is a rather broad feature towards the north-east with a position angle of 40$^\circ$, which coincides with the position angle of the galaxy's major axis.
There might be a counterpart on the south-western side,  whose existence is suggested by it kinematics. 
The reddening map shows that this  feature splits into two parallel arms accompanied by dust. Its connection with the centre resembles the base
of spiral arms rather than a jet. One also observes  here the close match of dust and gas. Its kinematic properties, however, suggest that it might be the debris of a galaxy-scale disc.

   The other radial feature points towards  the
    south-west  and is much
  fainter  (position angle  200$^\circ$). Its emission precisely traces   a straight line, starting from very near the centre (about 1\arcsec) and extending over 3.5 kpc. 
  
  There is no indication
that the  ambient  conditions for ionisation are different in any way (see Sect. \ref{sec:NII}). 
Since such a high degree of collimation  is not possible under stellar dynamical conditions with velocity dispersions about 200 km/s and differential rotation, this feature
suggests the presence of a jet.  There is no visible counter jet. Remarkably, there is no connected dust.
 One further observes that the 'jet' is nearly perpendicular
to the putative outflow described in Sect. \ref{sec:centre}. 
We come back to these features in Sect. \ref{sec:kinematics}.

\subsection{The NaI D map}

As a characterisation of the atomic gas component, we show a map of the interstellar absorption/emission of the NaI D line
The interstellar feature becomes visible after subtracting the stellar template spectra via the fitting with PyParadise. In fact, we fitted a Voronoi tessellation with an S/N that depends slightly on the MUSE pointing, typically
S/N=10.
 The stellar  NaI D is one of the strongest absorption lines in a late-type
spectrum. We therefore  did not expect the template subtraction to leave a residuum that really shows the interstellar absorption \citep{concas17}. Here we have  a lucky case where the highly structured dust provides proof for
the existence of the absorbing medium,
which would not be available  when  dealing with a more diffuse gas distribution. We note the amazing correspondence of structure in spite of the very different spatial resolution. To what level radial velocities and equivalent widths are numerically trustworthy  deserves more investigation and is beyond our scope. To our knowledge, this use of the NaI D-line in an early-type galaxy is the first in the literature. The absorption of resonant lines has been discussed extensively in the context of galactic winds (e.g. \citealt{veilleux05,rupke15,roberts19}
and references therein).  

NaI D  in emission indicates resonant scattering in an outflow \citep{prochaska11,rupke15}.  Resonant scattering is  expected to occur in regions of low dust absorption. Indeed, the emission appears outside the dust structure. Striking is the emission of the region at (650,680), where
there is [NII]-emission as well, but the dust structure shows a sharp cutoff.  
The overall picture resembles a cold outflow perpendicular to a disc (to compare with Fig. 1 in \citealt{roberts19}).  The sketch in Fig. \ref{fig:cartoon} illustrates this scenario.

\section{Gas kinematics}
\label{sec:kinematics}
The velocity maps for different lines reveal a wealth of details, the discussion of which must be postponed to a future paper with a catalogue character.
 Here we show the most interesting kinematic
details that are important for our conclusions. 

 
\subsection{The large scale}

\begin{figure}[th!]
\begin{center}

\includegraphics[width=0.9\textwidth]{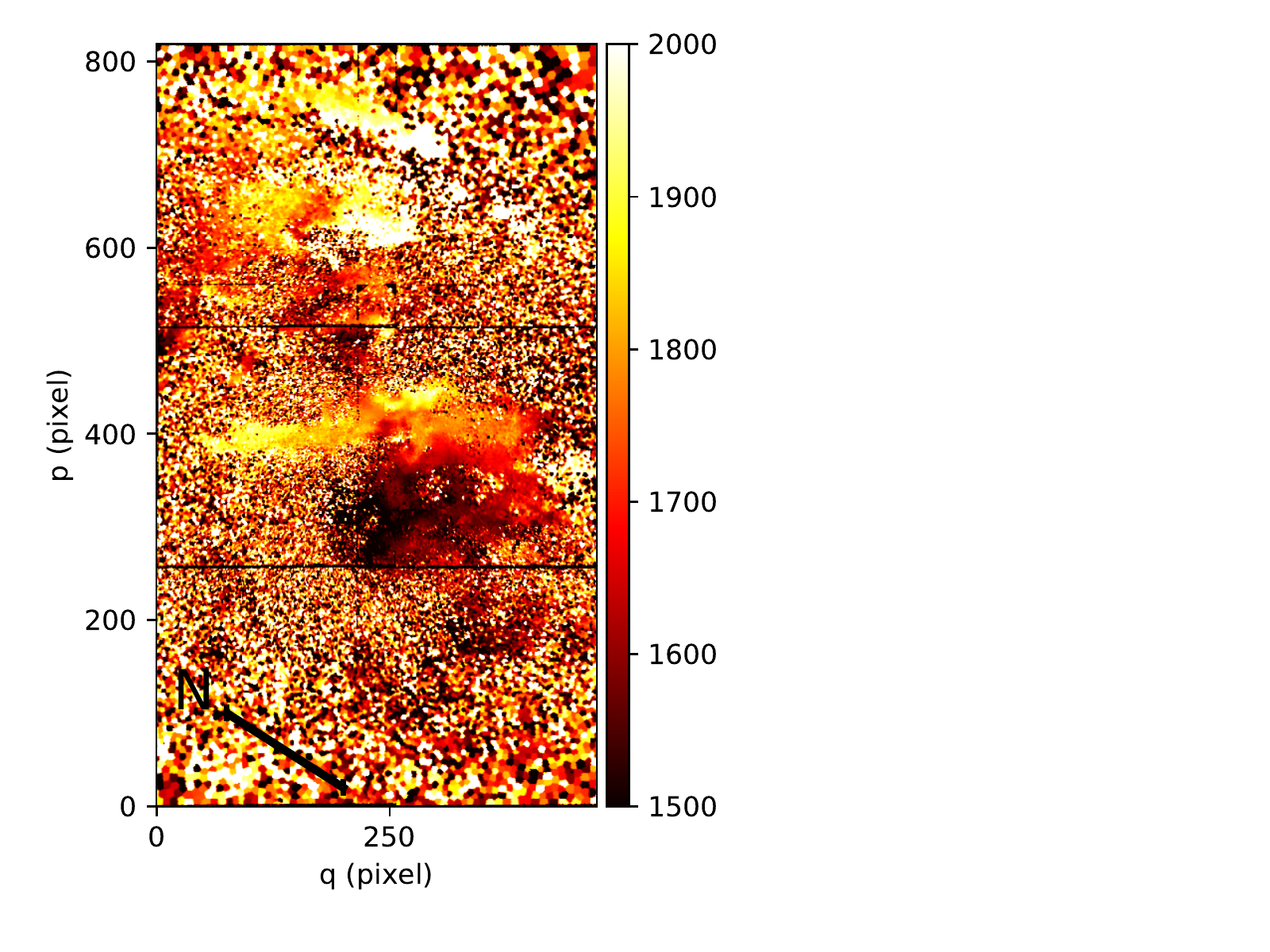} 
\caption{Velocity map of [NII] for the part of the  mosaic that contains [NII]-features.  The size is $80"\times164"$. North is indicated by the black line. The systemic velocity is 1750 km/s. The northern parts tend to have higher velocities and the
southern parts lower velocities, which fits to a general outflow. The most northern parts are the fastest, which may indicate a curved shape along the line-of-sight. This is less obvious in the
southern parts. Other small-scale velocity fields are superimposed. The horizontal structure covering the centre  is aligned with the galaxy's major axis and follows its rotational direction.
We therefore identify it with a dusty gas disc that is   frequently seen in early-type galaxies. 
 }
\label{fig:kinematics_all}
\end{center}
\end{figure}

Figure \ref{fig:kinematics_all} shows the velocity map of [NII]6583 for the inner parts of the MUSE mosaic, where [NII] is present. We tend to find higher velocities (with respect to the systemic velocities)  in the northern part and lower velocities in the southern part. The highest velocities are found in the most northern section,
which may be explained by an outflow that is curved along the line-of-sight. The highest velocities are about 2200 km/s. This pattern  is not obvious in the southern part. Velocities of this amount cannot be
of stellar dynamical origin, which provides additional support for an outflow hypothesis 
 Adopting an outflow velocity of 500 km/s, the corresponding dynamical age is about $10^7$y. The only other known object in NCG1316
that is this young is SH2 \citep{richtler12b}.
Velocity fields of a smaller scale  are superimposed. We show the central part with a higher resolution in the next section.


\subsection{A gaseous disc with  dust}
\label{sec:disc}
The radial structure on the [NII]-map that points towards the north-east (see Fig.\ref{fig:kinematics}), where the reddening map reveals a rather inconspicuous  double arm structure in the dust, seems to be of a different nature than the bulk of the gas or dust,
which defines an axis perpendicular to it.   The double arm structure can be seen as an  inclined disc or ring,
but there are more indications:
The direction coincides with the bulge major axis, and  the velocities match the projected bulge velocities in the central region but deviate outwards.  
This is expected for a rotating disc because the mean line-of-sight velocity of the rotating stellar bulge becomes lower than that of the rotating disc as its radius increases towards the north-east (and
higher as its radius increases towards the south-west). 
 The velocity structure is symmetric with respect to the centre. However, the south-western part of the disc is disturbed by the dust and gas,
probably coming from the centre. 

The inclination can only be determined  with some uncertainty because the disc itself is not well defined. We estimate the maximal north-east extension on the [NII]-map to be 30\arcsec\ and the width to be 6\arcsec, corresponding
to an inclination angle of 78$^\circ$. 
Trying to find the highest outermost measurable radial velocity of [NII]  on the north-eastern side, we find 1950$\pm$20 km/s at a projected distance of
3.3 kpc. We therefore have $v_{rot} = 200\pm20$ km/s, corresponding to $3.2\times10^{10} M_\odot$.  This is much lower than the expected value. Our luminosity model from \citet{richtler12a} demands $9.6\times10^{10} M_\odot$ with  $M/L_R$=2.5, which would be characteristic of an intermediate-age population. 
This conflict probably means that either the highest [NII]-velocity is not the circular velocity or that the true inclination is lower.  
Solving this disagreement will require a more detailed dynamical investigation, including an investigation of the stellar population. 
It is suggestive to consider this disc as a young version of the central dust lanes that are frequently observed in early-type galaxies. The mere existence of a disc requires that its age is at least a few orbital periods,
which at 3.3 kpc and with $3.2\times10^{10} M_\odot$  is  0.1 Gyr. Therefore, the disc must be much older than the perpendicular structure (see next section).

We note that an NaI D absorption is not visible, which is another difference. This may indicate a purely molecular gas disc. In the CO map of \citet{morokuma19}  (their Fig. 5), their 'extended' region  appears to have the correct position angle and
can perhaps be identified with the south-eastern part of our disc, but it appears to be too far to the south. The north-eastern part is not detected. The possibility remains that there is no cold gas at all and that the disc is completely ionised.

The disc that we describe here is not the
disc found by  \citet{schweizer80}, which rotates along the minor axis with a much higher velocity. We can only suspect that the general north-south velocity pattern created 
the impression of a rotating disc when observed with spectroscopic long slits.  Further discussion on this is presented in Sect. \ref{sec:comparison}.  
 
\begin{figure}[th!]
\begin{center}
\includegraphics[width=0.4\textwidth]{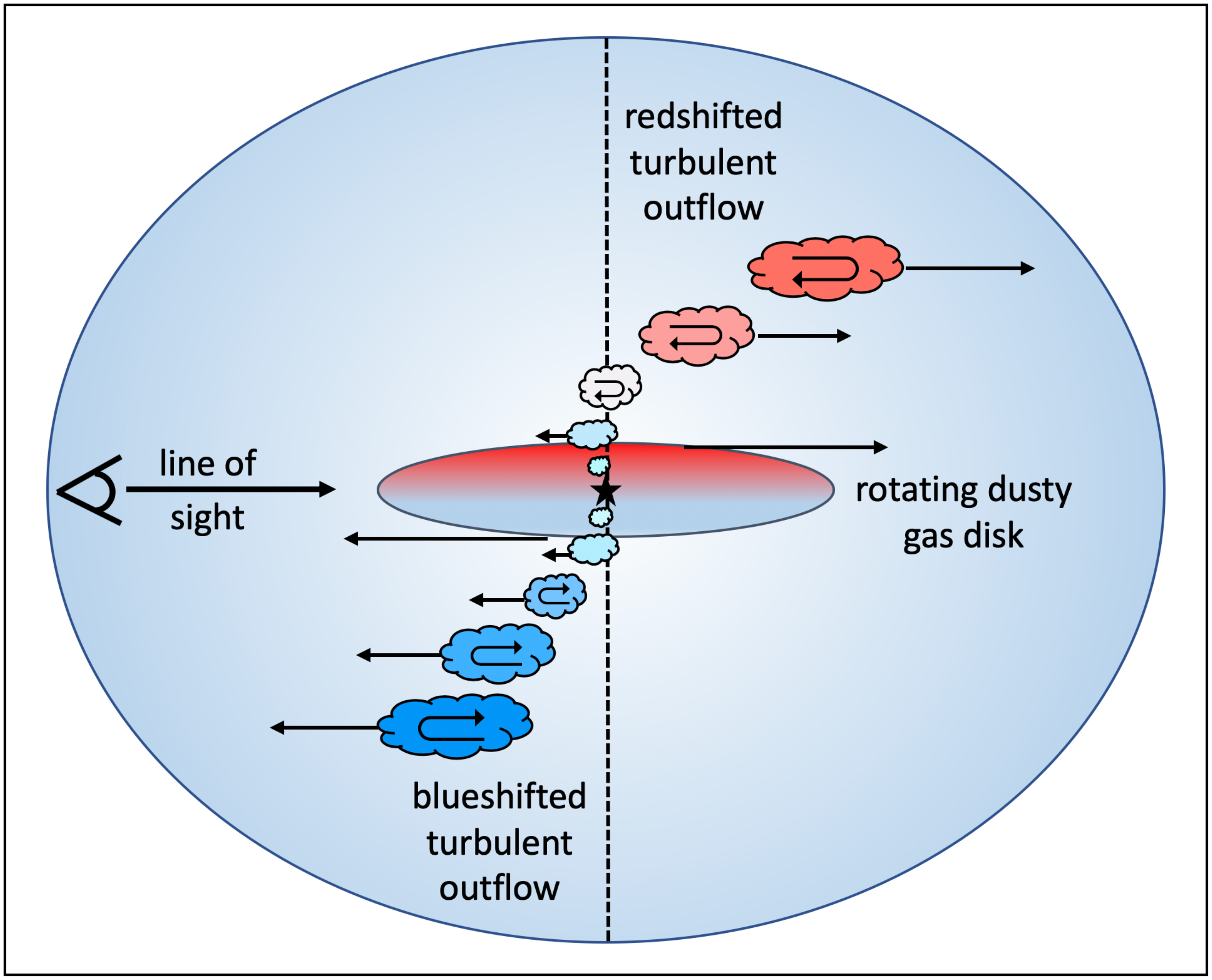} 
\caption{Illustration that visualises the observer's perception of a bipolar, turbulent, and curved outflow.  The blueshift in the south and the redshift in the north give the impression of a rotating disc seen edge-on.}
\label{fig:cartoon}
\end{center}
\end{figure}

\begin{figure}[]
\begin{center}
\includegraphics[width=0.5\textwidth]{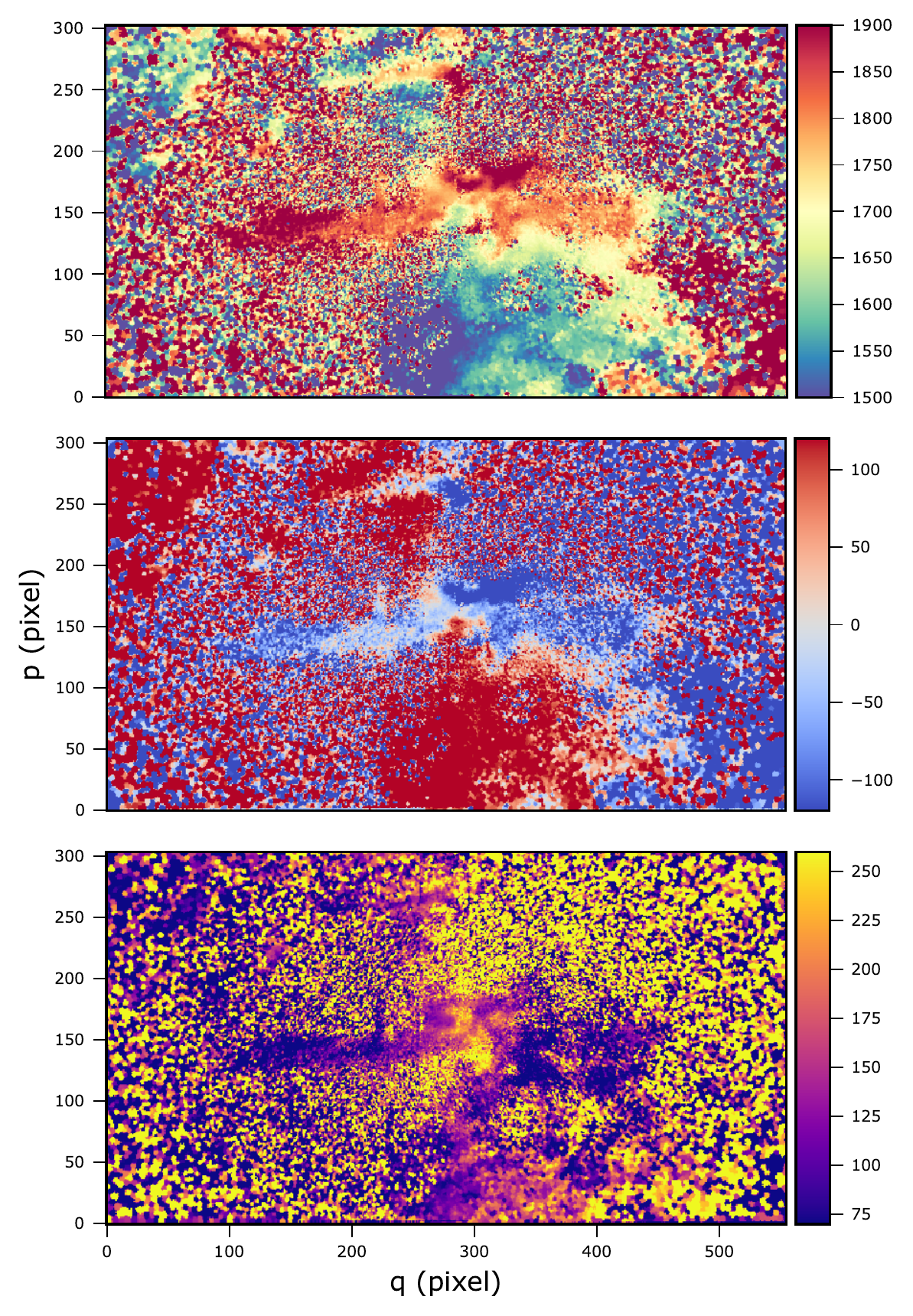} 
\caption{Kinematics of the central region of NGC 1316. The regions have sizes of $80"\times30"$. The orientation is the same as in Fig. \ref{fig:kinematics_all}. The nucleus is at p=138, q=326. {\bf Upper panel:} [NII]6583 velocities for the two central fields. 
 Conspicuous is the horizontal feature
with kinematical properties that resemble the bulge rotation. 
 {\bf Middle panel:} Map of the difference 'stellar velocities minus [NII]':  The horizontal feature
follows the direction of the bulge's rotation. This strongly suggests the existence of a dusty  gas
disc. {\bf Lower panel:} Velocity dispersion of the [NII]6583 line.
The mere dispersion values may hide complicated velocity structures changing on very small scales.   
 }
\label{fig:kinematics}
\end{center}
\end{figure}

 Figure \ref{fig:cartoon} illustrates our scenario as viewed  from the west onto an inclined dusty disc as well as a bipolar-structured turbulent outflow,
 which is curved. The observer looks from the 'left'  and sees in projection
 what may be misinterpreted  as an edge-on disc that is blueshifted in the south and redshifted in the north.

\subsection{Velocity dispersion and lifetimes} 

\begin{figure}[]
\begin{center}
\includegraphics[width=0.5\textwidth]{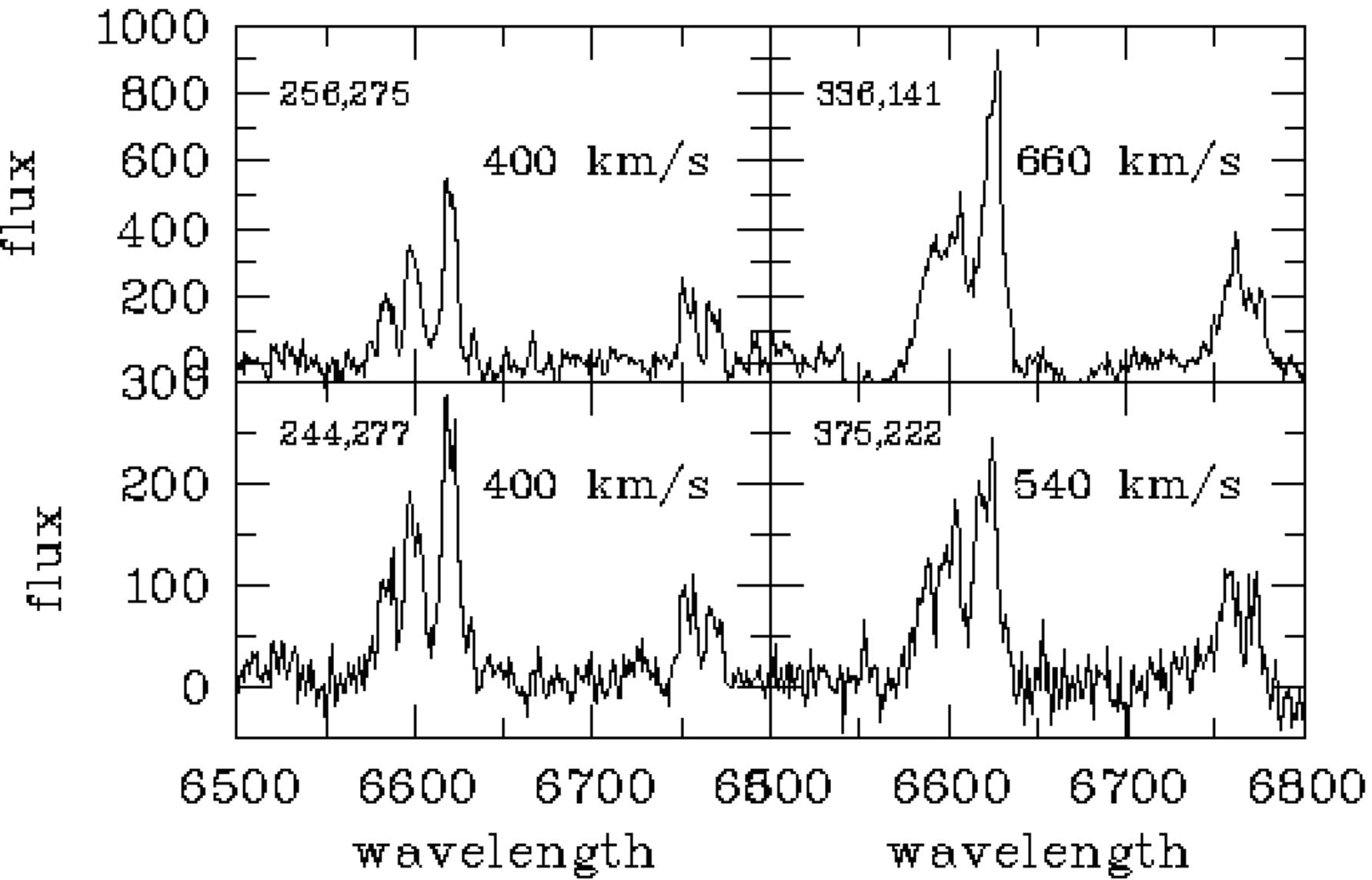} 
\caption{Four examples of  multi-component velocity fields in the emission lines after subtracting the model fits of the stellar population. The spectral range includes, from left to right:
[NII]6549, H$\alpha$,[NII]6583, [SII]6717, and [SII]6731. The pixel coordinates are given. The velocity dispersion values refer to [NII]6583 but should be equal for all lines. We suggest that such velocity fields emerge in outflows.   }
\label{fig:curlfields}
\end{center}
\end{figure}

For the evaluation of the velocity dispersion map of [NII], it is important to realise that the width of the [NII]-line, if fitted by a simple Gaussian,
cannot reproduce  the true velocity dispersion field that
is characterised by small-scale variations and multiple components. Figure \ref{fig:curlfields} gives an impression.
 The velocity dispersion can reach values higher than 200 km/s. These structures are, of course, 
not bound and will disperse rapidly.  An order-of-magnitude estimation of the dissolution time scale is $t_{diss} \approx R/\sigma$, with R being a characteristic radius and $\sigma$ the velocity dispersion.
With R= 500 pc and $\sigma$ = 200 km/s, $t_{diss}$ = 2.4$\times10^6$ yr.   This is about 40 times lower than the orbital period at a radius of 3.3 kpc. 

\subsection{Comparison with molecules}
\label{sec:comparison}
 The recent work of \citet{morokuma19}, using data from ALMA and Keck,  enables an illustrative  comparison of our results with the distribution of molecules and their kinematics.  \citet{morokuma19}   present intensity 
and kinematic maps of CO  with a spatial resolution of 16\arcsec, covering an area similar to our MUSE mosaic. Their  maps do not resolve the dust structures in detail, but they do identify four 'clumps' of CO emission, which 
closely resembles   a strongly blurred  dust map, including the dust structure located about 20\arcsec\ south-west of the centre.  A detailed comparison with our Fig. \ref{fig:kinematics_all} is not possible either, but the overall tendency of high velocities in the north and low velocities in the south
is reproduced. A long-slit spectrum, taken with the Low Resolution Imaging Spectrometer (LRIS) at the Keck telescope,  
provides better spatial resolution but less precise velocities along the 'dust axis', which is identified with Schweizer's disc.  It confirms what is known from the kinematics of the molecular gas. Our disc, described in Sect. \ref{sec:disc}, is not visible on the CO maps.  It is interesting that the velocity dispersion of CO is mostly very low, about 50 km/s, although their velocity resolution should
resolve the kinematic substructure, which is visible in the [NII]-lines. They conclude that the complexity of the velocity field does not support a simple rotation, and they favour a scenario where  individual movements of gas blobs
are superimposed onto a rotational movement.  Such a scenario is obviously not viable within our present   context. Our interpretation of the Morokuma-Matsui et al. observation is that it is a low-scale molecular outflow.
Molecular outflows have  been identified in many galaxies (see \citealt{lutz20} for recent work and literature compilation), with outflow rates of up to several hundred solar masses per year. A  rough estimation with 
an outflow mass of $3.6\times10^8 M_\odot$ (the Morokuma-Matsui et al. molecular mass), an outflow radius of 5 kpc, and an adopted outflow velocity of 500 km/s gives  588 $M_\odot$/year, applying Eq. (3) from \citet{lutz20}, which
assumes sphericity. The outflow in NGC 1316 happens in a narrow cone. Adopting $20^\circ$ for the opening angle, a factor of 0.01 gives about 5 $M_\odot$/year.      
   

\subsection{Comparison with HI}

After previous failed attempts to detect HI in the central region of NGC 1316 (e.g. \citealt{horellou01}), \citet{serra19} succeeded in finding central HI using observations with MeerKAT. 
The very centre still appears devoid of HI, but there are two separate HI-clouds very close to the centre that appear in Table 2 of \citet{serra19} as 'NGC 1316' and are quoted with
an HI-mass of $4.3\times10^7 M_\odot$. The exact coordinates of the two components are not given; therefore, we cannot give a convincing correspondence to features in [NII] or
NaD I. However,  the northern component has a higher radial velocity than the southern component, and they therefore  fit the general outflow scheme.
 
 Nevertheless, the very centre hosts atomic gas, as indicated by NaI D. An answer to the question of why this gas is not seen in HI is beyond the scope of this work.
 Without making
a strong claim, we suspect that the local velocity dispersion of HI is so high that the emission is diluted over too many MeerKAT channels and drops below the detection limit for
a single channel. For comparison, the [NII]-velocity dispersion in this region reaches values higher than 200 km/s. 

A further observation is that the disc is not seen in HI.\ This corroborates the assumption that it is also not visible in NaI D. 

\citet{serra19} find more  HI emission outside the central region.\ This, among other sources, confirms the presence of HI in the star-forming region SH2 \citep{schweizer80,richtler17}. 

\section{The central region}
\label{sec:centre}
\begin{figure}[]
\begin{center}
\includegraphics[width=0.5\textwidth]{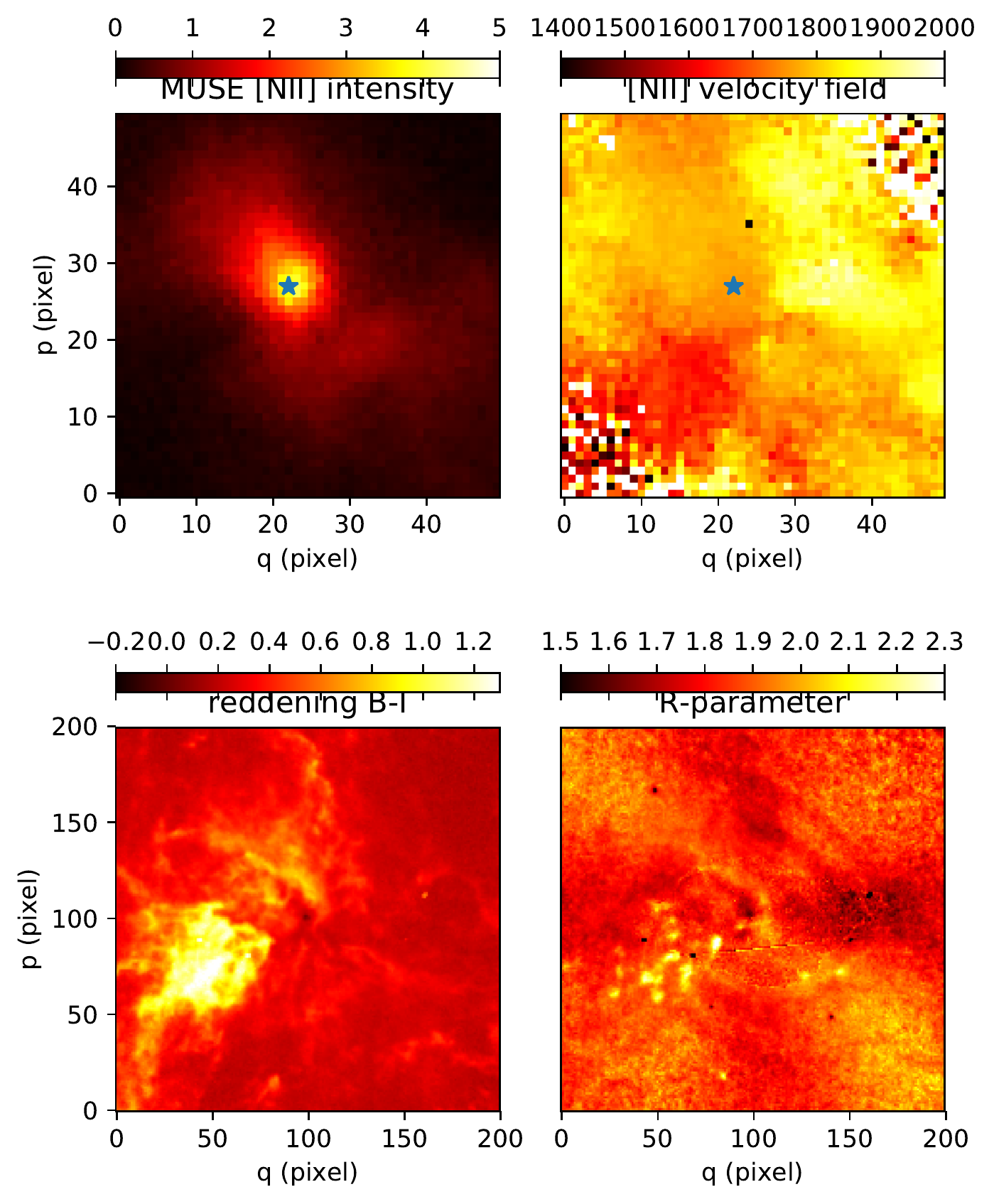} 
\caption{Central region of NGC 1316  under various aspects. The images cover 10$\times$10 $\rm arcsec^2$, corresponding to 863$\rm \times$863 $pc^2$.  The optical centre is indicated by green star symbols in the two upper panels.  {\bf Upper-left panel:}  [NII]-map.  Flux units are $\rm 10^{17} erg/sec/cm^2/\AA $. The [NII]-luminosity is
strongly concentrated in the centre, but the emission is asymmetric, with an extension towards the north-east. {\bf Upper-right panel:}   [NII]-velocity map. A remarkable structure
is visible, which we interpret as an outflow. The green star indicates that the 'kinematical centre' is offset from the optical centre
by about 1\arcsec (corresponding to 86 pc in projection). {\bf Lower-left panel:}  Reddening map in the very centre, which visualises the dust.  The nucleus is the blue point source at p=100, q=100. The dusty  cone-like 
 region coincides with the approaching part of the outflow, while no dust seems to be connected
with  the receding part; this may be explained by a faint background. {\bf Lower-right
panel:}  Map of the R-parameter from Eq. \ref{eq:reddeninglaw}.  A well-defined value is R=1.85. One observes tiny spots  indicating higher R-values and  higher absorption. We speculate that these spots may be
pockets of efficient grain growth.}
\label{fig:centre}
\end{center}
\end{figure}

\begin{figure}[]
\begin{center}
\includegraphics[width=0.4\textwidth]{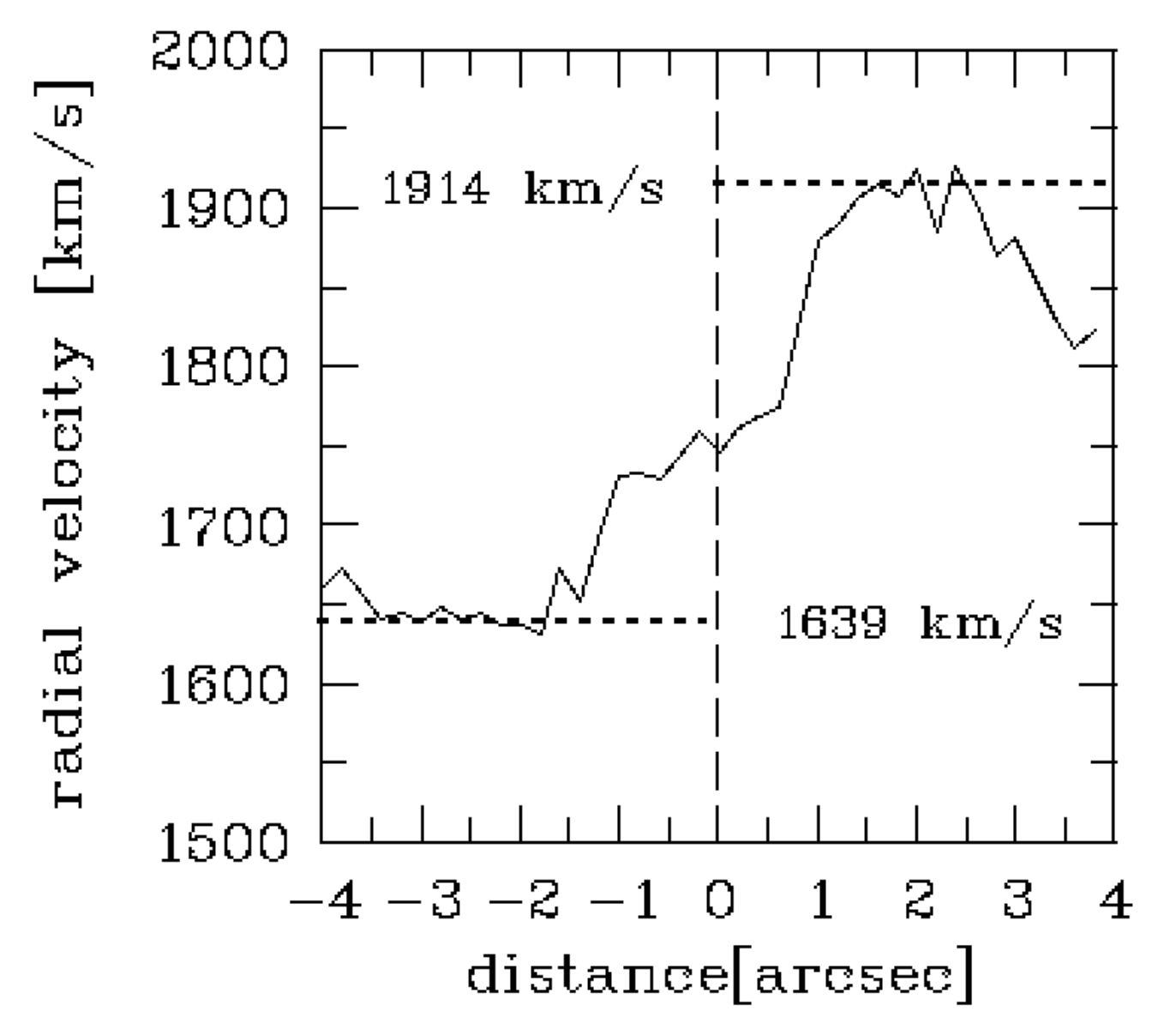} 
\caption{Extraction of the velocities along the cone-like velocity feature that is seen in the upper-right panel of Fig. \ref{fig:centre}. The zero-point is the kinematical centre. We interpret this feature
as an outflow.  }
\label{fig:outflow}
\end{center}
\end{figure}

The centre is a particularly interesting region. The nucleus (or what we think is the nucleus) itself is not hidden but appears in the optical as a  point source, slightly bluer than its  environment by $\Delta B-I \approx 0.2$. Deprojecting the spherical galaxy model of \citet{richtler14}, and adopting 5 pc as the emission radius, one needs about a factor of 30, or 3.7 mag, of brightening in B to achieve that
colour shift. This is not explainable by a population colour,  only by non-thermal emission, so one may speak of a weakly active nucleus.
The upper-left  panel of Fig. \ref{fig:centre} shows the central [NII]-emission. The star symbol indicates the location of the nucleus. Although the bulk of the [NII]-emission seems shifted towards
the north-east, the locations of the peak emission of [NII] and the optical are not distinguishable on the scale of the MUSE data.   The central [NII]-velocity map is shown in the 
upper-right panel. The cone-like velocity feature is striking, showing a low-velocity part on the south-eastern side and a high-velocity part on the north-western side. 
 Figure \ref{fig:outflow} shows an extraction of
velocities, using a 3 pixel-wide slit with a position angle of 315$^\circ$ passing through the kinematical centre to which the zero-point  corresponds. 
We interpret this feature as an outflow rather than a disc. The intensity map does not show a corresponding disc-like structure. 
One also observes more high-velocity gas nearby, which does  not correspond to a putative disc-like structure and
may come from previous outflows.  As seen in Fig. \ref{fig:outflow}, the radial velocity curve is not very symmetric with respect to the centre.
   
The kinematical centre is offset from the optical centre by about 1\arcsec\ to the south.\ This means, in the case of a disc, that the centre-of-mass would not coincide with the nuclear point source.

Only the low-velocity feature coincides  with the cone-like dust structure visible in the lower-left panel;  there is no correspondence with dust on the high-velocity side.  We suggest the explanation
that the low-velocity flow is directed towards the observer in the foreground of the centre, while the dust in the opposite direction does not have the bright background needed to be visible as an absorbing screen.


\section{The reddening law}
Interesting findings come from considering the reddening law of the dust in NGC1316.
The reddening law is written in the form
\begin{center}
\begin{equation}
A_{F435W} = R_{F435W}*E(B-I)
\label{eq:reddeninglaw}
.\end{equation}
\end{center}

Using the reddening map, we  constructed a map of  R-values  that, according to Eq. \ref{eq:reddeninglaw}, can 'correct'
for the dust, in the sense that multiplying this map with the original image allows the dust structure to disappear. It is not self-evident
that such a uniform  R-value exists, given the appreciable range of $R_V$-values in the Milky Way (e.g. \citealt{fitzpatrick99}), but it
does exist. A value of $R_{F435W} = 1.85$ almost completely removes  the dust structure.  
Given that R-values in the Milky Way fill
an interval 2.5 $<$ $\rm R_V$ $<$ 4, depending on the Galactic positions of reddening probes, it is striking that the dust is so homogenous.
The Galactic value for the F435W-filter is 1.6 \citep{fitzpatrick99}. 

 The absorption values are not very high, with maxima of about 0.8 mag. 
The corrected image was modelled with $\it ellipse$ under IRAF; the image was also divided by this to make small residuals visible. 
We calculated
a map

\begin{equation}
-2.5 \times log10(I_{F435,corrected}/I_{F435,model}) + 1.85
\label{rmap}
,\end{equation}

in which tiny residuals from the applied R-value also become visible. The central region of this map is shown in the lower-right panel of Fig. \ref{fig:centre}.
The elliptical model is not perfect but leaves some structural residuals, in particular the disc-like structure of the inner isophotes.

Although R=1.85 is precisely defined, there are small-scale spots where R-values are higher by a small amount. These spots are by no means artefacts: The
same structures emerge when other ACS filters are used (for the sake of brevity, we do not show this).
Moreover, the distribution of these little patches is not random: They prefer the highest local absorption. The lower-right panel of Fig. \ref{fig:centre}
shows the central region as  a particularly interesting zone. The small spots are clearly related to the dusty outflow.  The high number of
spots becomes understandable if the line-of-sight forms a small angle with the axis of the putative outflow and thus samples a large volume.
Larger R-values may indicate larger dust grains. In combination with the association of higher absorption, it is tempting to speculate that these spots
are pockets of higher gas density and a more effective dust formation by grain growth from the ISM (e.g. \citealt{zhukovska08,hirashita15}). 
We remark that elevated R-values do not necessarily mean larger dust grain sizes. One could also think of unresolved lines-of-sight
in which the absorption assumes very high values, for example in molecular cloud cores. This problem cannot, of course, be attacked
with our data; it needs ALMA observations.


\section{Summary and conclusions}
We present initial, intriguing results from analysing a mosaic of 12 MUSE fields of the inner region of NGC 1316 (Fornax A). 
As a pilot study, we concentrated on a few striking phenomena in combination with archival HST/ACS imaging.   
 
 Our main findings are as follows. The LINER-like line emission with the [NII]6583 \AA\ as the strongest line exhibits similar line-strength ratios independent
from the local line luminosity, which strongly suggests  post-AGB stars to be the sources of ionisation. 
Line emission is strongly concentrated in the centre. 
The fainter line emission at larger radii
matches the shape of the dust features in detail, indicating that there is little ionised gas outside
the dust pattern.

The global [NII]-velocity pattern is best understood by a galaxy-wide outflow where at least the northern part is curved along the line-of-sight. Near the centre, the outflow vector is almost tangential,
becoming more radial (i.e. having higher radial velocities) at larger distances.

  We generated a map of the interstellar  NaI D absorption that emerges after modelling and the subtraction of the stellar galaxy light. Stunningly, the neutral gas distribution also follows many dust patches in detail.
  Methodologically, our study shows, for the first time, that  NaI residuals of early-type stellar populations are indeed useful for detecting cold gas. 
We also see NaI D emission at the northern and southern  borders of the dust pattern, illustrating models of resonant line emission in a galactic wind.
The close  match of both ionised gas and neutral gas with dust strongly argues for the fact that dust and gas were never separated  and, thus, for a common origin inside a small volume. 
This must be the nuclear region. 
 The velocity dispersion of the
gas  tracing the dust structures indicates a short lifetime; therefore, conventionally discussed dust formation scenarios in AGB stars and SNe are not viable. Instead, we propose dust formation in a
nuclear outflow.

However, this is valid only for the dust distribution that defines an axis perpendicular to the major axis of the bulge. A second dust component, striking in an [NII]-map as a radial feature, defines an axis along the
major axis of the bulge.  [NII] shows a low-velocity dispersion and a velocity pattern that agrees with the bulge's rotation. Interestingly, no NaI D absorption is visible.
We interpret this feature as a central disc-like structure analogous to the central dust lanes in elliptical galaxies.

A straight radial [NII]-feature of a projected length of about 4 kpc, not connected with dust,  seems to trace a jet that itself is not visible in the optical.
The [NII]-velocity map of the central region indeed shows a bimodal outflow pattern with a projected velocity of approximately 200 km/s, which coincides with  a central dust structure. From HST/ACS images, we constructed a map of the R-parameter in the
reddening law.  The reddening  is very homogeneous, in contrast to the Galactic law, but shows tiny patches with R-values that are slightly higher than the environmental R-values. They
are not visible on the reddening map. We suggest that these
features are unresolved molecular cloud cores with high absorption.  
 We note that apparent stellar sodium abundances and sodium abundance gradients may be questioned in the presence of interstellar atomic sodium \citep{sarzi18}.

If dust production is a part of nuclear activity, one would expect to see this in other galaxies as well. The first morphological test is to use high resolution archival images to check whether a dust pattern or filament 
is connected with the nucleus. This is an easy task. We encourage others to do this test with, for example, NGC 1022, 1459, 1700, 4150, 4589, and 5102. The last galaxy is a particularly 
convincing case (Richtler et al. in preparation). Even the small dust filament that emanates
from the nucleus of M87   has apparently remained unnoticed despite the prominence of the nucleus \citep{sparks93}.



 \begin{acknowledgements}
We thank the anonymous referee for a constructive report. Without Bernd Husemann and his help to use PyParadise, this paper 
would not have been written.  TR acknowledges support  from  the BASAL Centro de
 Astrof\'{\i}sica y Tecnologias
Afines (CATA) PFB-06/2007.  This work has emerged from TR's visits at ESO/Garching under the  ESO science visitor programme.   We thank Jakob Walcher for giving access to the reduced MUSE data prior to
the advent of ESO phase 3 products.     TR thanks the Astronomisches Institut der Ruhr-Universit\"at Bochum for hospitality and computer time.
TR thanks Johanna Hartke for discussions and help with matplotlib. 
This research made use of Montage. It is funded by the National Science Foundation under Grant Number ACI-1440620, and was previously funded by the National Aeronautics and Space Administration's Earth Science Technology Office, Computation Technologies Project, under Cooperative Agreement Number NCC5-626 between NASA and the California Institute of Technology.
\end{acknowledgements}

\bibliographystyle{aa}
\bibliography{MUSE.bib} 
  
\end{document}